# Challenges in the crystal growth of Li$_2$FeSiO$_4$

Waldemar Hergett[a], Christoph Neef[a,b], Hans-Peter Meyer[c], Rudiger Klingeler[a,d]

[a]Kirchhoff Institute of Physics, Heidelberg University, 69120 Heidelberg, Germany
[b]Fraunhofer Institute for Systems and Innovation Research ISI, 76139 Karlsruhe, Germany
[c]Institute of Earth Sciences, Heidelberg University, 69120 Heidelberg, Germany
[d]Centre for Advanced Materials, Heidelberg University, 69120 Heidelberg, Germany

Abstract

The high-pressure optical floating-zone method enables single crystal growth of the *Pmnb* high-temperature phase of Li$_2$FeSiO$_4$. The influence of growth conditions on crystal quality, phase homogeneity, and impurity formation in Li$_2$FeSiO$_4$ is studied. The use of different starting materials, i.e., either the *P*12$_1$/*n*1 or the *Pmn*2$_1$ polymorph, as well as optimization of various growth conditions is investigated. Several *mm*$^3$-sized high-quality single crystals are obtained by the choice of the *Pmn*2$_1$ polymorph as the starting material. A general challenge of Li$_2$FeSiO$_4$ crystal growth is polymorph control during crystallization. While the temperature gradient at the solid-liquid interface seems to have significant impact on stabilizing the *Pmnb* high-temperature phase, growth velocity has no evident effect.

1. Introduction

The orthosilicate Li$_2$FeSiO$_4$ is intensively studied as a new-generation cathode material for lithium-ion batteries, as it features favorable characteristics, such as environmental harmlessness, thermal stability, good operation voltage, and high theoretical capacity [1, 2]. A main issue for long-term cyclability of Li$_2$FeSiO$_4$ is its rich temperature-dependent polymorphism including energetically close structures *Pmn*2$_1$, *P*12$_1$/*n*1, and *Pmnb*, which are stabilized at ∼ 400 °C, ∼ 700 °C, or ∼ 900 °C, respectively [3, 4, 5, 6, 7, 8, 9]. Several methods of preparing polycrystalline samples of the desired polymorphs have been reported, and a variety of investigations on the structural, morphological, and electrochemical properties of Li$_2$FeSiO$_4$ have been performed.

In this work, we present a systematic investigation of the influence of growth conditions on crystal quality, phase homogeneity, and impurity formation in Li$_2$FeSiO$_4$. We report the growth of *mm*$^3$-sized *Pmnb*-Li$_2$FeSiO$_4$ single crystals by the optical floating-zone (FZ) method at an elevated Ar pressure. In particular, two different synthesis routes yielding *Pmn*2$_1$ and *P*12$_1$/*n*1 starting materials were examined in regard to applicability for a successful FZ growth of Li$_2$FeSiO$_4$ single crystals. We discuss the role of the starting polymorphs in determining which phases and grain morphologies are formed during the FZ process.

2. Experimental procedure

Raw materials for the feed rods were prepared by two different synthesis routes. A one-step synthesis method was applied in the route *A*, yielding the *Pmn*2$_1$ polymorph: FeC$_2$O$_4$·2H$_2$O with 5% excess above stoichiometric requirements, Li$_2$CO$_3$ and SiO$_2$ were mixed in a planetary ball mill with acetone, dried and heated at 370 °C for 12 h. The resultant product was reground, pressed into pellets and sintered at 800 °C for 6 h using a ramp of 300 °C/h on both heating and cooling. The tube furnace was operated at 100 mbar with constant Ar flux of 250 standard cubic centimeters per minute and pressurized to 1400 mbar under static atmosphere after heating up to 430 °C in order to avoid vaporization of lithium oxide. An observed loss on ignition is expected due to dehydration and decomposition of $FeC_2O_4 \cdot 2H_2O$ and $Li_2CO_3$ and it is consistent with the removal of the gaseous products (CO, $CO_2$, $H_2O$). For synthesis route *B*, $Li_2SiO_3$ was prepared from $Li_2CO_3$ and $SiO_2$ by conventional solid-state reaction. Subsequently, $Li_2SiO_3$, metallic Fe and $Fe_3O_4$ in molar proportions of 4:1:1 were ball-milled with acetone, dried and sintered at 850 °C for 6 h, which resulted in *P*12$_1$/*n*1. Ramping and atmosphere conditions were chosen as in route *A*. In both cases, reground powders were pressed to rods with diameters of 6 mm and typical lengths of 70 - 110 mm under an isostatic pressure of 60 MPa. Immediately after pressing, the rods were transferred to an argon-filled glovebox for storage. The rods were dense and ready to be used as feed and seed rods without additional sintering. All crystals were grown in a high-pressure floating-zone (FZ) furnace (HKZ, SciDre) [10]. The entire crystal growth procedure was monitored with a high-resolution CCD camera equipped with specialized filters and lenses. In-situ temperature measurements were obtained by a two-color pyrometer, applying the stroboscopic method. Polarized-light and scanning electron microscopy (SEM) was used for analyzing the microstructure and crystal quality. EDX microanalysis was conducted by means of a Leo 440 scanning electron microscope equipped with an Inca X-Max 80 detector (Oxford Instruments). Powder X-ray diffraction (XRD) patterns of the as-prepared samples and reground crystals were measured on a Bruker D8 Advance ECO diffractometer equipped with an SSD-160 line-detector in Bragg-Brentano geometry using CuK$\alpha_{1,2}$ radiation. All the XRD patterns were analyzed by employing the Rietveld refinement with the FullProf 2.0 software [11]. X-ray Laue back-scattering method was utilized to determine the crystal orientation.

3. Results and discussion

XRD patterns of the starting materials obtained via the two routes described above are displayed in Fig. 1. Both diffractograms indicate good crystallinity of the pristine materials. For the route *A* material, in addition to the *Pmn*2$_1$ polymorph of $Li_2FeSiO_4$ the data show presence of a small amount of metallic α-Fe which originates from decomposition of excessive $FeC_2O_4 \cdot 2H_2O$. Rietveld analysis reveals the presence of 97(2) % of *Pmn*2$_1$-$Li_2FeSiO_4$ and of 2.5(4) % of α-Fe, respectively. For the route *B* material, only the *P*12$_1$/*n*1 polymorph can be identified. Thus, depending on the synthesis conditions and the choice of precursors, different $Li_2FeSiO_4$ polymorphs are obtained. Results of the feed preparation process are summarized in Tab. 1.

Table 1: Process description for route A and B feed preparation.

| | Route A | Route B |
|---|---|---|
| Reagents | $FeC_2O_4 \cdot 2H_2O$, $Li_2CO_3$, $SiO_2$ | $Li_2CO_3$, $SiO_2$, $Fe_3O_4$, Fe Intermediate: $Li_2SiO_3$ |
| Remarks | 5 % excess of $FeC_2O_4 \cdot 2H_2O$ | Stoichiometric proportions |
| Process conditions | First step: 370 °C for 12 h Second step: 800 °C for 6 h | 850 °C for 6 h |
| Products | $Pmn2_1$-$Li_2FeSiO_4$ and α-Fe | $P12_1/n1$-$Li_2FeSiO_4$ |

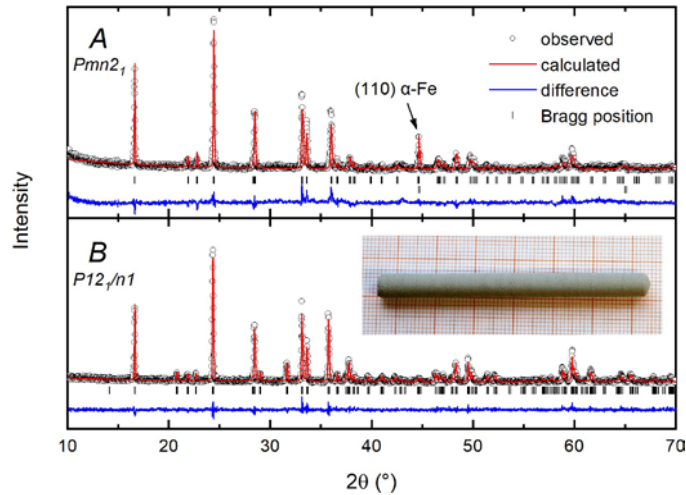

Figure 1: Rietveld-refined XRD patterns of the $Pmn2_1$ (route A) and $P12_1/n1$ (route B) polymorphs used as starting materials for the crystal growth. For the route A, the second row of vertical ticks indicates Bragg positions for the additional $Im\bar{3}m$ phase of metallic α-Fe. The peak marked with an arrow is attributed to the (110) lattice plane of α-Fe. Inset: picture of a typical route B feed rod used for the FZ process.

The crystal growth employing polycrystalline rods both from routes A and B was carried out in a high purity Ar atmosphere at an elevated pressure of 30 bar to suppress vaporization of $Li_2O$, and with a constant Ar gas flow through the growth chamber at a rate of 0.02 l/min. The feed and seed rods were counter-rotated for a better homogenization of the molten zone at a rate of 21 and 17 rpm, respectively. A number of FZ experiments with different growth rates ranging from 1.5 to 30 mm/h were conducted. Long-term stable growth was realized when the average temperature of the zone was held at ~1310 °C for route A and ~1290 °C for route B feed rods. Characteristic temperature profiles measured along the rods' axes by moving the pyrometer vertically in steps of 0.5 mm are shown in Fig. 2. For both starting materials, the curves exhibit a bell-like shape with a broad central region and a plateau-like shoulder at $z \approx 15$ mm. The shoulder is related to the liquid-solid coexistence region which reflects the melting temperature $T_m$. Stoichiometrically-synthesized material obtained through route *B* exhibits $T_m \sim 1190(10)$ °C, whereas, material from route *A* shows $T_m \sim 1240(5)$ °C, which is attributed to small amounts of excess Fe. A steep temperature gradient of ~150 °C/mm due to the stated growth atmosphere and sharp focusing of the light was realized close to the growth interface for all experiments. In general, large temperature gradients may be favorable for incongruently melting materials, as gentle temperature gradients may facilitate the melt to attack the feed rod and spill over the crystal [12]. Accordingly, we observe clear borders between the molten zone, on the one hand, and both the feed rod and the grown boule, on the other hand. No spills of the melt were encountered during the growth process (see inset photograph in Fig. 2).

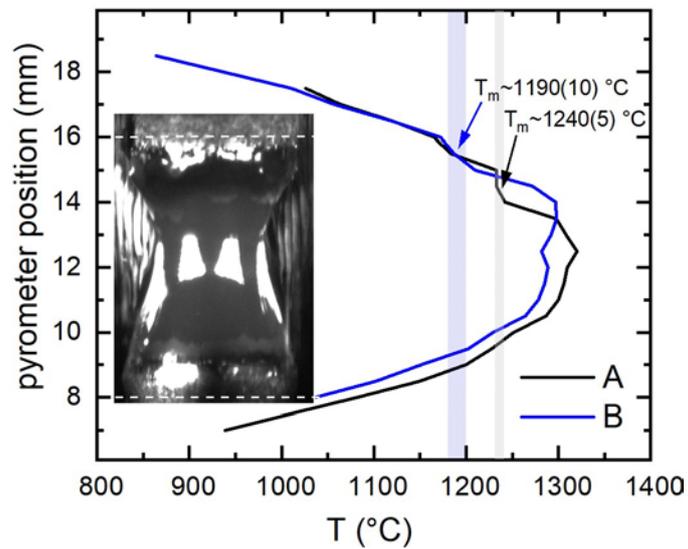

Figure 2: Temperature profiles recorded along the vertical axis of the feed rods from routes *A* (black line) and *B* (blue line) in the region of the molten zone. The shoulders show the melt-solid coexistence and enable reading off the melting temperatures $T_m$. The inset photograph shows a typical route *A* molten zone during the crystal growth.

The XRD patterns of the powdered crystals (output from the FZ process) indicate phase-pure materials only in the specimen obtained via route *A* [13]. In this case, Rietveld refinement of the experimental data in Fig. 3 shows that all observed diffraction peaks originate from the *Pmnb* phase of $Li_2FeSiO_4$. Whereas, route *B* yields the $P12_1/n1$ phase of $Li_2FeSiO_4$ along with 10(1)% impurities of $Cmc2_1$-$Li_2SiO_3$. The enhanced intensity of (020) (*Pmnb*) and (101) ($P12_1/n1$) peaks in Fig. 3 indicates preferred orientation of the crystallites within the powder, which was considered in the refinement procedure. We attribute this to the platelet-like shape of the crystallites (Fig. 3) originating from anisotropic cleavage properties (see, e.g., Refs. [14, 15]). More vigorous grinding did not reduce the preferred orientation effect. At room temperature, the refined lattice parameters are determined to be a = 6.280(5) Å, b = 10.652(9) Å, c = 5.037(3) Å for the *Pmnb*-polymorph and a = 8.228(7) Å, b = 5.019(5) Å, c =

8.231(6) Å for the $P12_1/n1$ polymorph, respectively. The values of refined lattice parameters agree well with those reported in the literature [5].

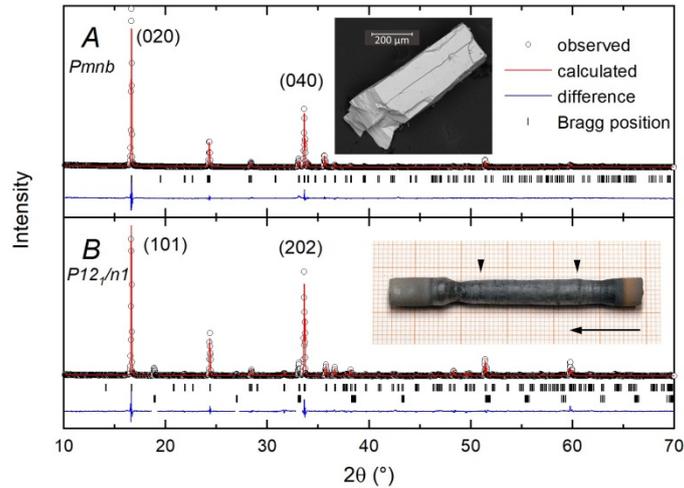

Figure 3: Rietveld-refined XRD patterns collected from powdered single crystals obtained via the routes *A* and *B*. Vertical ticks designate the Bragg positions of the main phases and, for *B*, of the $Cmc2_1$-structure attributed to $Li_2SiO_3$. Upper panel inset: SEM image of a *Pmnb* cleaved grain. Lower panel inset: as-grown boule from the route *B* experiment with the arrow indicating the growth direction. Black triangles mark approximately the beginning and the end of crystallization.

Notably, while the FZ process implies complete melting of the starting materials and very similar growth conditions, the XRD patterns show that crystals obtained from the routes *A* and *B* exhibit different crystal structures, i.e., *Pnmb* and $P12_1/n1$, respectively. Both structures, i.e., the *Pmnb* structure from route *A* as well as the $P12_1/n1$ structure from route *B*, are high-temperature phases of $Li_2FeSiO_4$ [3] which are stabilized at ~900 °C and ~700 °C, respectively. According to differential scanning calorimetry data, the structural phase transition *Pmnb* → $P12_1/n1$ appears upon slow cooling at ~724 °C [3]. We conclude that, despite slow growth velocity, the cooling process is too fast to allow this transition in our TSFZ-experiment. Changing the growth rate between 1.5 and 30 mm/h does not influence polymorphism of the obtained crystals. Instead, our data imply that in the complex interplay of nucleation, growth kinetics, and thermodynamics, the choice of the starting material is crucial for determining the formed phase(s). Tentatively, we assume that subtle differences in the oxygen stoichiometry of the starting materials are one of the most critical parameters that control which polymorph is formed during the FZ process. In addition, we suppose that the steep temperature gradient plays a significant role in stabilization of the *Pmnb* phase.

Energy-dispersive X-ray spectroscopy (EDX) was employed to determine the elemental compositions at various positions of different boules (Figs. 4 and 5). Specimens from routes *A* and *B* show clear differences in the microstructure and phase purity. In particular, while route *B* crystals suffer from a variety of cracks there are much larger single-crystalline grains in boules from route *A*. Several initial attempts using a stoichiometric rod composition for the route *A* failed to produce reasonably large crystals at any pulling rate between 1.5 and 30 mm/h while the formation of an Li$_2$SiO$_3$ impurity phase was observed. These issues were solved by introducing a small excess of Fe which enables the growth of single-crystalline specimens of *Pmnb*-Li$_2$FeSiO$_4$ at a comparatively fast growth rate of 10 mm/h. To be more specific, the optimal overstoichiometric conditions were identified by gradually increasing the surplus amount of FeC$_2$O$_4$·2H$_2$O in the initial synthesis step of the powers which were used to produce the feedrods and by investigating the purity of the boules resulting from the FZ process. Li$_2$SiO$_3$ impurities in the grown crystals were eliminated when a 5 % excess of FeC$_2$O$_4$·2H$_2$O was employed. A typical grain from the route *A* material is seen in Fig. 4a. The dominant phase (light grey) in Fig. 4a corresponds to the main phase of Li$_2$FeSiO$_4$. In addition, we observe white inclusions with dendritic morphology mainly in the periphery of the boule, suggestive of constitutional supercooling (CSC). We associate them with iron oxide, as indicated by the EDX data. The results obtained from the EDX analysis performed on selected representative spots and areas (mapping) are summarized in Tab. 2. The stated errors represent standard deviations of the averaged measurements. For the analysis, we balanced the positive charges of the cationic sites by an appropriate quantity of oxygen assuming a nominal oxidation state of +2 for Fe. Due to the high atomic weight difference between Si and Fe, and due to the lack of reference standards with a known Si:Fe ratio, the quantitative reliability of EDX analysis is limited. But the data are still sufficient to estimate stoichiometries and thereby identify chemical phases. EDX point scans, e.g., at c1 (see Fig. 4a), confirm a Si:Fe ratio which is very close to that of the nominal stoichiometry of Li$_2$FeSiO$_4$. In the dendrites (spot c2), our data confirm the absence of Si, while the amount of oxygen is not reliably quantified by the EDX microanalysis. This minor (dendritic) phase only accounts for a small fraction of the sample, and as such, it may be below the detection limit of XRD, explaining the absence of additional peaks in the XRD pattern. The (crystalline) structure of this phase remains unknown. Dark grey spots in the rim region are presumably caused by pores and microbubbles. Area scans (e.g., area c3 in Fig. 4a) do not show any compositional discrepancy compared to the main phase.

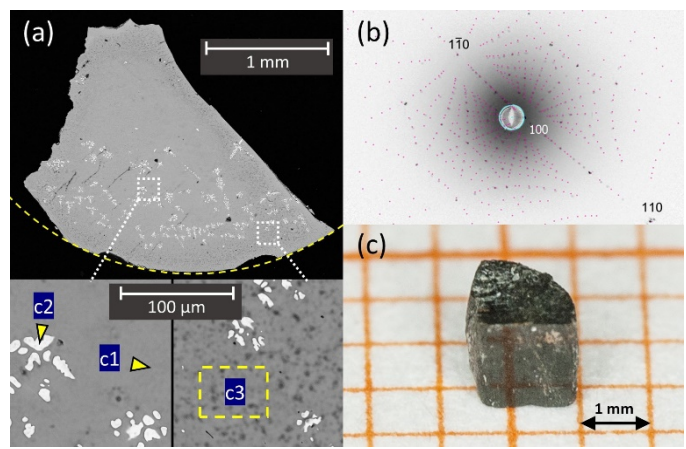

Figure 4: (a) SEM image of a *Pmnb*-Li$_2$FeSiO$_4$ grain obtained by the synthesis route *A* with local enlarged views of the dendritic FeO$_{1-\delta}$ inclusions. EDX analysis points and areas are marked and labelled c1 to c3. The dashed curve indicates the circumference of the boule. (b) Experimental (black spots) and simulated (purple spots) Laue diffraction patterns from the (100) face. (c) Photograph of an oriented *Pmnb*-Li$_2$FeSiO$_4$ single crystal.

*Pmnb*-Li$_2$FeSiO$_4$ forms air- and moisture-stable, translucent dark-brown crystals. Fig. 4c shows an exemplary single crystal of several mm$^3$ in volume. High crystallinity of this sample is illustrated by the Laue pattern in Fig. 4b taken perpendicularly to the (100) face. It is concluded that the dominant growth direction is in the *bc*-plane, i.e., perpendicular to the *a*-axis. Magnetization, thermal expansion and magnetostriction data assert high quality of the crystals [16].

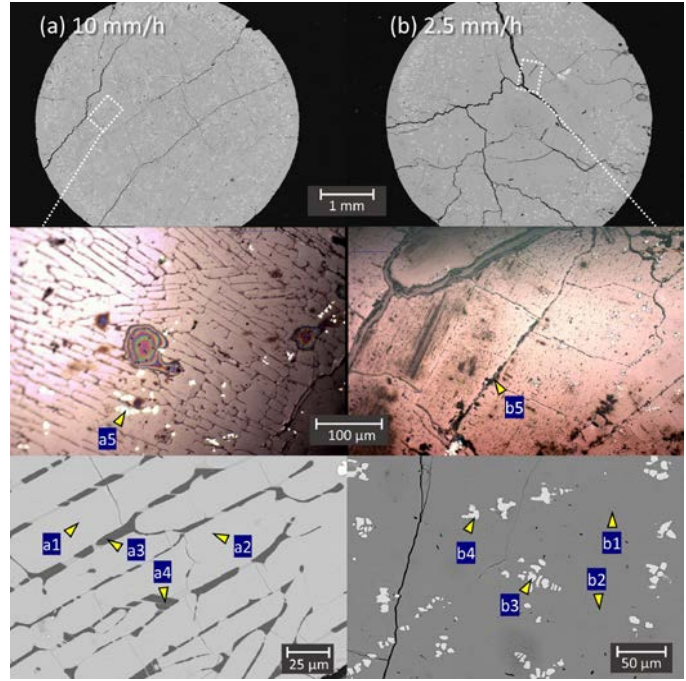

Figure 5: Upper panel: SEM images of the as-grown boules from route *B* with respective pulling rates of 10 mm/h (a) und 2.5 mm/h (b). Middle panel: enlarged polarized-light images of the marked areas. EDX measurement points are indicated by arrows (cf. Tab. 2). White spots are the minor impurity phases of FeO$_{1-\delta}$. Lower panel: Enlarged SEM images with marked spots for the EDX analysis.

Fig. 5 shows cross-sectional views obtained by SEM and polarized optical microscopy of the crystal boules grown at 2.5 mm/h and 10 mm/h from the route *B* material. Both crystal boules feature cracks propagating parallel to the growth direction, which are significantly more pronounced for pulling rate of 2.5 mm/h. In addition to the *P*12$_1$/*n*1-Li$_2$FeSiO$_4$ and Li$_2$SiO$_3$ components identified by XRD data, EDX measurements reveal the presence of dendritic FeO$_{1-\delta}$ similar to the route *A* results. We note that seeding from *Pmnb*-structured crystal boules did not significantly affect the resultant phase composition. Also, the use of above-stoichiometric Fe in the synthesis of route *B* materials did not improve the crystal quality, but in that case small amounts of the *Pmnb* phase were identified by XRD. The boule pulled at 10 mm/h (Fig. 5a) features eutectic solidification with very fine lamellae of an iron-free silicate phase (spots a3 and a4). The lamellae are embedded in a matrix with a close-to-stoichiometric ratio of Fe and Si (spots a1 and a2 in Fig. 5a). The data hence strongly suggest the lamellae being Li$_2$SiO$_3$ embedded in a matrix of Li$_2$FeSiO$_4$. The Li$_2$SiO$_3$ lamellae are generally well aligned along the rod axis with quite a regular spacing of 10 – 15 μm.

Table 2: Compositional EDX analysis of various points and areas as indicated in Figs. 4 and 5.

|  | SiO$_2$ (mol%) | FeO (mol%) |
|---|---|---|
| main phase (a1, a2, b1, b2, c1, c3) | 0.52(1) | 0.48(1) |
| FeO$_{1-\delta}$ impurities (a5, b3, b4, c2) | 0.003(1) | 0.997(5) |
| SiO$_{1-\delta}$ impurities (a3, a4, b5) | 0.98(1) | 0.02(1) |

At a growth rate of 2.5 mm/h, crack formation is more pronounced than at 10 mm/h (Fig. 5). As the solidification velocity decreases, lamellae are eliminated and, instead, the crystal seems to adopt a cellular morphology with a larger phase separation and a primary cell spacing in the range of 100 – 200 µm. Compositional analysis (spot b5) confirms enrichment of Li$_2$SiO$_3$ in the intercellular zone. Hence, the extent of Li$_2$FeSiO$_4$-Li$_2$SiO$_3$ phase segregation in the investigated system seems to be crucially driven by the solidification time. Crack generation can be attributed to stress caused by differential contraction between the two separated phases during cooling, since the phases' respective thermal expansion coefficients differ. Therefore, the higher degree of phase separation at 2.5 mm/h promotes more severe crack formation. A significant separation effect is also seen for the impurity phase of FeO$_{1-\delta}$ (spots b3 and b4) which develops dendritic structures in the Li$_2$FeSiO$_4$ matrix (spots b1 and b2) especially in the periphery of the boule. Since the powder XRD patterns do not reveal any signatures of crystalline iron oxides, the exact composition of this minor phase remains unclear. Optimized growth parameters and key results are summarized in Tab. 3.

Table 3: Summary of growth parameters and main characteristics for the two examined synthesis routes.

|  | Route *A* | Route *B* |
|---|---|---|
| Feed rod composition | *Pmn*2$_1$-Li$_2$FeSiO$_4$ and α-Fe | *P*12$_1$/*n*1-Li$_2$FeSiO$_4$ |
| Growth atmosphere | Ar: 30 bar, 0.02 l/m gas flow | |
| Melting temperature | 1240(5) °C | 1190(10) °C |
| Temperature of the stable molten zone | ~ 1310 °C | ~ 1290 °C |
| Pulling rate | 10 mm/h | 2.5 and 10 mm/h |
| Crystal composition | *Pmnb*-Li$_2$FeSiO$_4$ | *P*12$_1$/*n*1-Li$_2$FeSiO$_4$ and *Cmc*2$_1$-Li$_2$SiO$_3$ impurity phase |

4. Conclusions

The high-pressure optical floating-zone method enables single crystal growth of the high-temperature phases of $Li_2FeSiO_4$, namely *Pmnb* and *P*12$_1$/*n*1. Choice of the *Pmn*2$_1$ polymorph as the starting material yields $mm^3$-sized high-quality single crystals of the *Pmnb* structure at a pulling rate of 10 mm/h and an Ar pressure of 30 bar. By means of above-stoichiometric addition of Fe to the starting material undesired solidification of the $Li_2SiO_3$ phase was suppressed. A minor phase (≲ 1%) of Fe-rich dendritic precipitates is not accessible by XRD analysis. The high temperature gradient of ~150 °C/mm at the solid-liquid interface seems to control the formation of the *Pmnb* polymorph, while the growth velocity was found to have no evident effect. Choice of the *P*12$_1$/*n*1 polymorph as the starting material yielded *P*12$_1$/*n*1-structured single crystals. Due to the time-dependent solidification of the $Li_2SiO_3$ impurity phase and associated cracking, size of the crystal grains is limited to 100 – 200 μm. Access to $Li_2FeSiO_4$ single crystals facilitates further studies on anisotropic ionic transport and delithiation dynamics and can provide important insights into Li-diffusion mechanisms. Beyond that the high-pressure optical floating-zone method opens up new prospects for crystal growth of other members of this class of cathode materials, i.e., $Li_2MSiO_4$ (M = Mn, Ni, Co).

Acknowledgments

The authors thank Ilse Glass and Dr. Alexander Varychev for technical support. Financial support by the German-Egyptian Research Fund (GERF IV) through project 01DH17036 and by the Deutsche Forschungsgemeinschaft (DFG) through project KL1824/5 is gratefully acknowledged.

References

[1]   A. Nyten, A. Abouimrane, M. Armand, T. Gustafsson, J. O. Thomas, J. Electrochem. Commun., 7 (2005), pp. 156-160
[2]   Z. Gong, Y. Yang, Energy Environ. Sci., 4 (2011), pp. 3223-3242
[3]   G. Mali, C. Sirisopanaporn, C. Masquelier, D. Hanzel, R. Dominko, Chem. Mater., 23 (2011), pp. 2735-2744
[4]   S. Nishimura, S. Hayase, R. Kanno, M. Yashima, N. Nakayama, A. Yamada, J. Am. Chem. Soc., 130 (2008), pp. 13212-13213
[5]   A. Boulineau, C. Sirisopanaporn, R. Dominko, A. R. Armstrong, P. G. Bruce, C. Masquelier, Dalton Trans., 39 (2010), pp. 6310-6316
[6]   [6] A. R. Armstrong, N. Kuganathan, M. S. Islam, P. G. Bruce, J. Am. Chem. Soc., 133 (2011), pp. 13031-13035
[7]   T. Kojima, A. Kojima, T. Miyuki, Y. Okuyama, T. Sakai, J. Electrochem. Soc., 158 (2011), pp. A1340-A1346
[8]   C. Sirisopanaporn, C. Masquelier, P. G. Bruce, A. R. Armstrong, R. Dominko, J. Am. Chem. Soc., 133 (2011), pp. 1263-1265
[9]   M. Bini, S. Ferrari, C. Ferrara, M. C. Mozzati, D. Capsoni, A. J. Pell, G. Pintacuda, P. Canton, P. Mustarelli, Nature Sci. Rep., 3 (2013), 3452

[10]  C. Neef, H. Wadepohl, H.-P. Meyer, R. Klingeler, J. Cryst. Growth, 462, (2017), pp. 50-59

[11]  J. Rodriguez-Carvajal, Phys. B, 192 (1993), pp. 55-69

[12]  T. Ito, T. Ushiyama, Y. Yanagisawa, Y. Tomioka, I. Shindo, A.Yanase, J. Cryst. Growth, 363 (2013), pp. 264-269

[13]  W. Hergett, C. Neef, H. Wadepohl, H.-P. Meyer, M.M. Abdel-Hafiez, C. Ritter, E. Thauer, R. Klingeler, J. Cryst. Growth, 515 (2019), pp. 37-43.

[14]  W. A. Dollase, J. Appl. Crystallogr., 19 (1986), pp. 267-272

[15]  C. Neef, C. Jahne, H.-P. Meyer, R. Klingeler, Langmuir 29 (2013), pp.¨ 8054-8060

[16]  W. Hergett, M. Jonak, J. Werner, F. Billert, S. Sauerland, C. Koo, C. Neef, R. Klingeler, J. Mag. Mag. Mat., 477 (2019), pp. 1-3